\documentclass[
    ,final            % use final for the camera ready runs
  ]
  {aipproc}

\usepackage{epsfig}
\layoutstyle{8x11double}

\begin{document}

\title{Quasi-Periodic Oscillations from Low-mass X-Ray 
Binaries with Neutron Stars}

\author{Jean Swank}{
  address={Goddard Space Flight Center, Greenbelt, MD 20740}
}

%%\author{<author2>}{
%%  address={<common address for author2 and author3>}
%%}

%%\author{<author3>}{
%%  address={<common address for author2 and author3>}
%%  ,altaddress={<author1 address>} % additional visiting address
%%}

\begin{abstract}

Before the launch of the Rossi X-ray Timing Explorer ({\em RXTE}) it was
recognized that neutron star accretion disks could extend inward to
very near the neutron star surface, and thus be governed by
millisecond timescales. Previous missions lacked the sensitivity to
detect them. The kilohertz quasi-periodic oscillations (QPO) that {\em RXTE}
discovered are often, but not always, evident in the X-ray flux. In 8
years {\em RXTE} has found kilohertz signals in about a fourth of 100
low-mass X-ray binaries (LMXB) containing neutron stars.  The observed power
spectra have simple dominant features, the two kilohertz oscillations,
a low frequency oscillation, and band-limited white noise. They vary
systematically with changes in other source properties and offer the
possibility of comparison with model predictions. New information from
the millisecond pulsars resolves some questions about the relations of
the QPO and the spin.  Coherence, energy spectrum and time lag
measurements have indicated systematic behaviors, which should
constrain mechanisms.

\end{abstract}

\maketitle

%%%%%%%%%%%%%%%%%%%%%%%%%%%%%%%%%%%%%%%%%%%%
%% MAINMATTER
%%%%%%%%%%%%%%%%%%%%%%%%%%%%%%%%%%%%%%%%%%%%

\section{A Brief History of LMXB QPO}

Soon after the discoveries of Sco X--1 and Cyg X--2, it was realized
that accretion onto a neutron star in a binary was a likely source of
the X-ray emission. But while clear pulsations were seen in the flux
from Hex X--1, these sources exhibited no periodic signal. The
possibility was raised that accretion over a long lifetime had spun up
the neutron star to frequencies higher than could have been measured
in the early observations \citep{alpar82}. Successive missions strove 
to increase
their sensitivity to higher frequencies. {\em EXOSAT} and {\em Ginga}
pushed the frontier to about 200 Hz.  The world before {\em RXTE} is below
200 Hz.

{\em EXOSAT} discovered timing signals, but quasi-periodic signals
rather than the coherent clock of the neutron star. QPO were found in
many X-ray sources in the galactic bulge. The frequencies varied in the
1-50 Hz range. {\em EXOSAT} proportional counter data provided spectral
information at the same time. 
\citet{HasvdK89} showed that the spectral variations fell in two categories,
denoted ``Z'' and ``Atoll'' and that the QPO frequencies 
depended on the source's position in a plot of hard versus soft ``colors'' or 
energy ratios. 
The widths and amplitudes varied also in systematic
ways. 
\citet{Wijnands01} compiled the {\em RXTE} version of a figure summarizing the
properties of both Z and Atoll sources. 

At first the frequency appeared to be positively correlated with
the X-ray luminosity and a simple explanation  was attractive,
the magnetic beat frequency model \citep{alpar85}. 
The accretion rate through the
disc, should be stopped eventually, by a magnetosphere due to the
neutron star, but closer to the neutron star because the magnetic
field was much weaker. The Kepler period of gas in the disk would beat
with the spin frequency of the neutron star to cause brightness
oscillations. Changing the accretion rate would change the
magnetospheric radius, the Kepler frequency at the boundary, and thus
the beat frequency. It implied spin rates of 50-350 Hz in several cases
\citep{GL92}.

However the model was not a satisfactory fit to the data from several
sources and there was evidence that the luminosity
was not a good measure of the accretion rate. The character of
bursts and their recurrence rate changed in 4U 1636--53 as it moved
through the ``Atoll'' pattern \citep{vdK90},
while the luminosity did not increase smoothly. 
In Cyg X--2 \citep{Has90,Vrtilek90} and Sco X--1 \citep{Vrtilek91} 
UV emission 
decreased as the X-ray flux increased, while the magnetospheric beat
frequency model implied it should increase \citep{Has90}. 
Nevertheless coherent
oscillations were sought
\citep{Vaughan94} and upper limits of less than 0.5 \% were achieved for
frequencies below 200 Hz.

The idea that the magnetic fields of the neutrons stars are $10^{2} - 10^{4}$ 
lower than the $10^{12}$ G of ``classical'' pulsars 
was advanced to explain the failure to detect strongly channeled accretion 
flow that should show up as pulsations and the 
higher frequencies.

\begin{figure}

\resizebox{1.0\textwidth}{!}{
  \begin{tabular}{lr}
   \epsfig{file=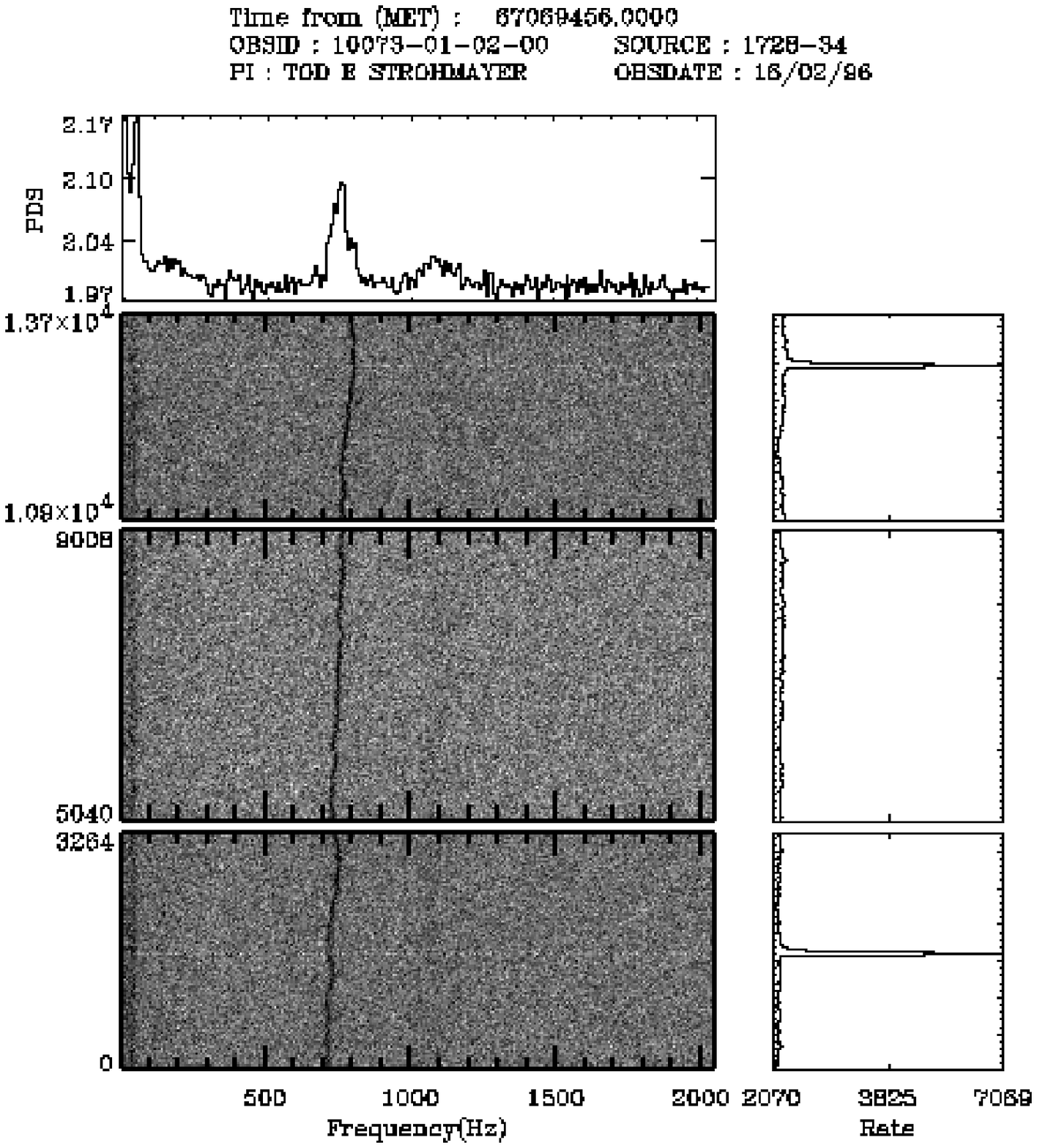,width=2.9in} & 
   \epsfig{file=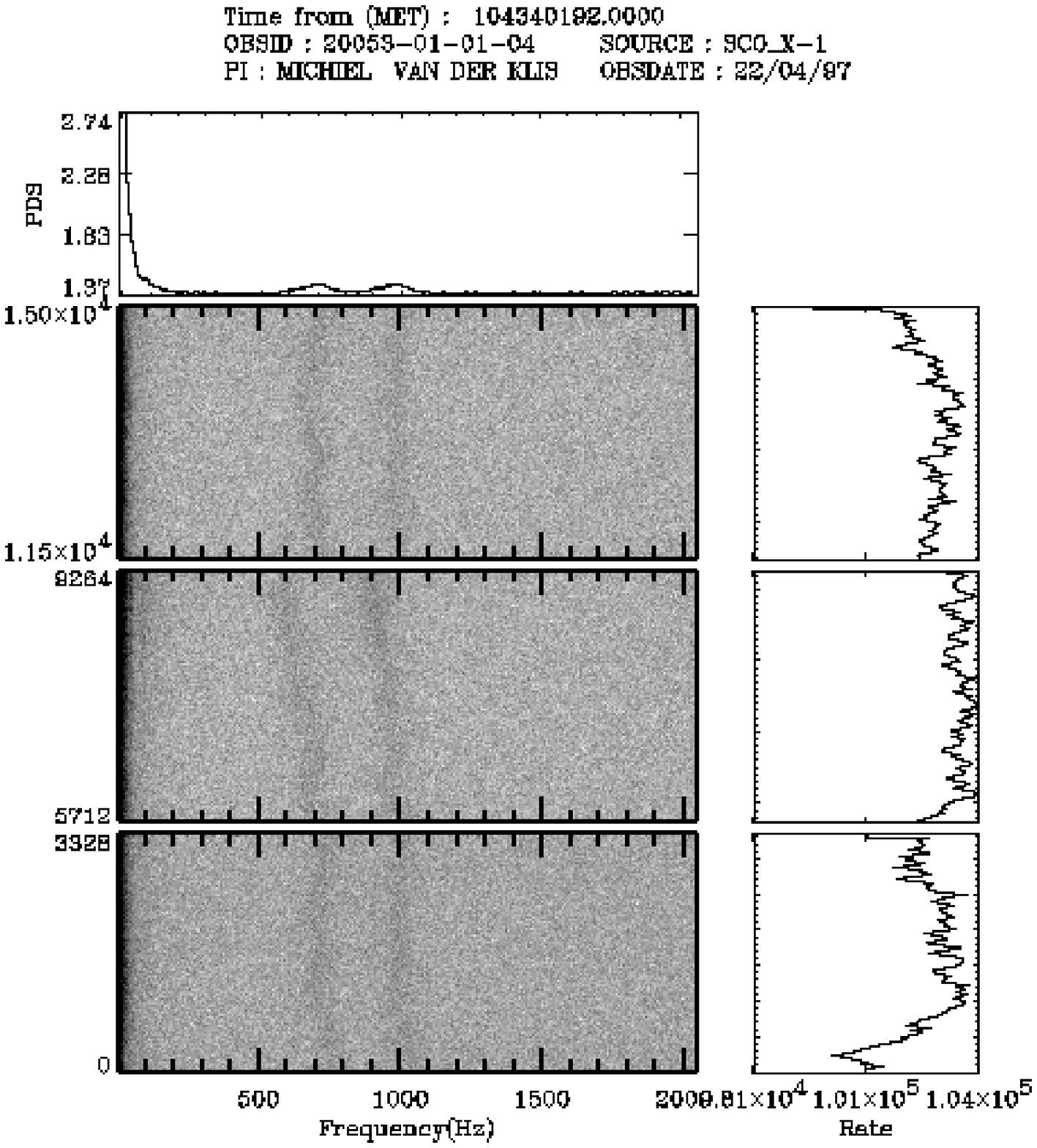,width=2.9in} \\
  \end{tabular}
}

\caption{Dynamical Power Spectra of 4U 1728-34 and of Sco X-1.
For each the spectra are shown for three consecutive observation intervals. 
At the top is an averaged Fourier power spectrum for all the intervals. 
For each source on the right is the light curve. The intervals included
two bursts from 4U~1728--34.
}
\label{fig:1}
\end{figure}

Kluzniak and Wagoner saw that accretion disks around low magnetic
field neutron stars could be very interesting  if the
equation of state of nuclear matter meant that neutron stars were
inside the innermost stable orbit of orbiting material. The accretion
disk could extend down to the innermost stable orbit and be truncated
there rather than at the magnetosphere. A signal might even indicate
the Kepler frequency of the inner most stable orbit
\citep{Kluz85,Kluz90}.
%(Kluzniak \& Wagoner 1985; Kluzniak, Michelson, Wagoner \& 1990). 
These papers foresaw that signals bearing the imprint
of General Relativity could come from these sources.

In 1996, {\em RXTE} began observing and the first observations of the Atoll
source 4U 1728--34 by \citet{Stroh96} and the Z source Sco X-1 by
\citet{vdK96} showed signals with frequencies in the range that orbits
close to neutron stars would have. Figure 1 shows several of the
important aspects of the kilohertz QPO discovered in the flux from 21
LMXB. As the count rates vary, the QPO center
frequencies vary significantly compared to the widths of the features.
The features in the Atoll and the Z source are very similar. The
phenomena and the physical models that have been explored during the 8
yr since the discovery have been described in several review articles 
\citep[See][]{vdK00,Wijnands01}.

Looking back at why {\em RXTE} could detect the signals while previous
missions did not, the increase in sensitivity came from several
factors.
The number of sigmas of the detection of a QPO feature can be expressed as 

\begin{equation}
n_{\sigma} = (1/2) S^{2}/(S+B)(rms/S)^{2} \surd (T/\Delta \nu)
\end{equation}

\noindent Here, S is the source count rate, B the background rate, rms the root
mean square variance in S, T the duration of the observation and
$\Delta \nu$ the width of the QPO feature. This scales as the detector
area. The PCA has observed with a maximum of 6250 cm$^2$, compared to
{\em EXOSAT}'s 1600 cm$^2$, but {\em Ginga} had 4000 cm$^2$, and did not
detect kiloHertz oscillations because of insufficient time resolution. Other 
factors - background, noise, dead time, low duty cycle of observations - 
have influenced the sensitivity to these phenomena. So far, {\em RXTE} 
has been the only 
instrument to detect them.

\section{Observed Characteristics of the Two KHz QPO}

\subsection{Frequency Range}

Low-mass X-ray binaries have a wide range in X-ray luminosity, from
apparently exceeding the Eddington limit for a neutron star of 1.4 M
to 0.5 \% of it. Yet for sources at both extremes the frequencies
observed for the upper of the two kilohertz oscillations range from
approximately 300 Hz to 1100 Hz. This was apparent early in the
exploration of QPOs and remains true now
\citep{SSZ98,vdK00}. (The highest frequency, 
although only 2.6 $\sigma$,  is 1330 Hz
for 4U 0614+09 \citep{vdK00,Straaten00}).
\citet{Zhang97}
%Zhang et al. (1997)
deduced from this that the frequency must depend only on
properties of the neutron star, independent of the mass accretion
rate. It could either be the radius of the neutron star or the radius
of the inner-most stable orbit (ISCO). It seemed more likely to
be the ISCO than the radius, in that surface  behavior would be
more likely to depend on the accretion rate. \citet{Kaaret97} also
pointed out that if the ISCO was responsible for the peak frequency,
it was a test of General Relativity. As the number of sources
accumulated, \citet{Ford00} exhibited that this independence of
the frequency range on the luminosity continued to hold, using fits to the 
simultaneous spectral data to determine more accurate luminosities.

Figure 2 shows that there is a slight trend for the
lower luminosity bursters to exhibit highest frequencies a little
higher than those of the Z sources at high luminosity. Interestingly,
none of the bright Atoll sources (e.g. GX 3+1, GX 9+9, GX 9+1, GX 13+1)
have shown oscillations. They fill in
the luminosity range between the brightest of the bursters, 4U
1820--30 and the Z sources.

\begin{figure}
\resizebox{1.0\columnwidth}{!}
%{\includegraphics{../figs/M\'{e}ndez-Ford.ps}}
{\includegraphics{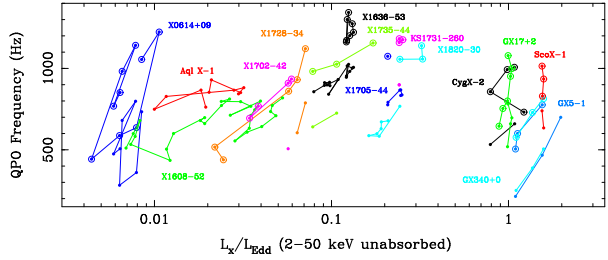}}
\caption{Frequencies of upper kilohertz oscillations for LMXB of the range
of luminosities. These values are for a subset of all the observations. 
The highest frequency points for 4U 1636-53 are of disputed significance.}
\label{fig:2}
\source{Ford et al. 2000}
\end{figure}

We now believe we know the rotation period P for 16 of the LMXB which
either have coherent oscillations or have oscillations during bursts,
For some of these we have a good estimate of the distance and
therefore the X-ray luminosity, L$_X$. The accretion rate onto the
neutron star could be through the disk or from a corona, so that for
the accretion rate in the disk, $dM/dt \leq
L_X/(GM/R)$.  If we suppose, as did \citet{White97}, that adiabatic
evolution keeps the neutron star rotation period approximately in
equilibrium with an average L$_X$, this is related to P and $\mu = B
R^3$. Table 1 gives values of B and the corotation radius R for 4
burst sources and also for 5 Z sources (for M=1.4 M$_{\odot}$ and R =
10 km). 

We don't have any burst oscillations for a Z source. But, the spin
period is believed to be within 15 \% of either the difference between
the two kilohertz frequencies, or twice it. The values of $\Delta \nu$
are relatively high, so that twice it would make these rates the
highest.  The values for B and the corotation radius R$_{co}$ in Table
1 assume a spin rate $\nu_s \approx \Delta \nu$. (If $\nu_s$ is twice  that, B
would be 2.25 times lower, and R$_{co}$ 1.6 times lower.) In this simple
treatment the magnetospheric radius is $\propto
L_{X}^{-2/7}$. Radiation drag and magnetospheric reaction are
neglected and approximate values  used, so that the estimates
can only be expected to hold to about a factor of 2.  

The resulting values for B imply that for the Atoll bursters B is a
few times $10^8$ Gauss, and for the Z sources a few times $10^9$
Gauss.  For both cases R$_{co}$ is about 2-3 times R$_{ISCO}$ = 6 G
M/c$^2$ =12.5 (M/1.4M$_{\odot}$) km, neglecting relatively small
relativistic corrections. A picture in which the disk extends close to
the ISCO and approaches it according to some scaling of the luminosity
appears quantitatively justified.

\begin{table}
%\begin{tabular}{lrrrrrr}
\begin{tabular}{lrrrr}
\hline 
\tablehead{1}{l}{b}{Source} &
  \tablehead{1}{c}{b}{Hz} &
% \tablehead{1}{c}{b}{P(ms)} &
  \tablehead{1}{c}{b}{L$_{37}$\citep{Ford00}} &
  \tablehead{1}{c}{b}{B($10^8$G)} &
% \tablehead{1}{c}{b}{R$_{co}$} &
  \tablehead{1}{c}{b}{R$_{co}$/R$_{isco}$}\\
\hline
4U 1728--34 & 363 & 1.0 & 3 & 2.6\\
4U 1636--53 & 581 & 3.2 & 3 & 1.9\\ 
KS 1731--26 & 524 & 6.0 & 5 & 2.0\\
4U 1702--43 & 330 & 1.2 & 4 & 2.7\\ 
Sco X--1 & 307 & 40 & 25 & 1.2\\
GX 5--1 & 298 & 37 & 25 & 2.9\\
GX 17+2 & 294 & 22 & 19 & 3.0\\
Cyg X--2 & 346 & 29 & 17 & 2.7\\
GX 340+0 & 339 & 30 & 19 & 2.7\\ 
%4U 1728--34 & 363 & 2.8 & 1.0 & 3 & 3.2 & 2.6\\
%4U 1636--53 & 581 & 1.7 & 3.2 & 3 & 2.4 & 1.9\\
%KS 1731--26 & 524 & 1.9 & 6.0 & 5 & 2.5 & 2.0\\
%4U 1702--43 & 330 & 3.0 & 1.2 & 4 & 3.4 & 2.7\\
%Sco X--1 & 307 & 3.3 & 40 & 25 & 3.6 & 1.2\\
%GX 5--1 & 298 & 3.4 & 37 & 25 & 3.7 & 2.9\\
%GX 17+2 & 294 & 3.4 & 22 & 19 & 3.7 & 3.0\\
%Cyg X--2 & 346 & 2.9 & 29 & 17 & 3.3 & 2.7\\
%GX 340+0 & 339 & 3.0 & 30 & 19 & 3.4 & 2.7\\
\hline
\end{tabular}
\caption{LMXB parameters assuming Spin Equilibrium}
\label{tab:a}
\end{table}

\subsection{Size of the Emission Region}

Models that have been advanced to explain the oscillations do not
agree on the mechanism for producing them. It is not in the scope of
this paper to delve into them. But if the disk extends inside a radius
2-3 times the ISCO, the inner rim of the disk as well as the surface
of the neutron star could be a source of X-rays. 

A statistically demanding measurements is that of time delays
of photons of different energy bands. This has been done for data in
which a QPO is very strong, in particular the lower of the two
frequencies, for 4U~1636--53, 4U~1608--52, 4U~1828--34, and 4U~1702--429. 
The result was initially surprising.

In the case of the black hole Cygnus X-1, in the low hard state hard
photons are delayed behind soft photons and it has been understood in
terms of a coronal model, in which a corona of high temperature
electrons is cooled by Compton scattering low energy photons
originating in the disk. In order to explain some parts of the energy
spectra of the low mass X-ray binaries, Comptonization off a small
corona has been discussed. Some of the same elements undoubtedly
should be in both Cyg X-1 and the low mass X-ray binaries with neutron
stars.

The sign of the delay for the QPO in several sources is the
opposite. This suggests the intrinsic properties of the emission
regions are more likely to cause the time delay than is
Comptonization. Figure 3 shows the delay as a function of energy for
4U~1636--536 and 4U~1608--522 \citep{Kaaret99}.  Converting the time
delay to a light travel time, the contributions can be no more than 20
km apart. This would be consistent with emission from the neutron star
surface and from the inner edge, or from different parts of the disk.
Such measurements would benefit from a higher throughput instrument.

\begin{figure}
\resizebox{1.\columnwidth}{!}
%{\includegraphics{../figs/pdelay.ps}}
{\includegraphics{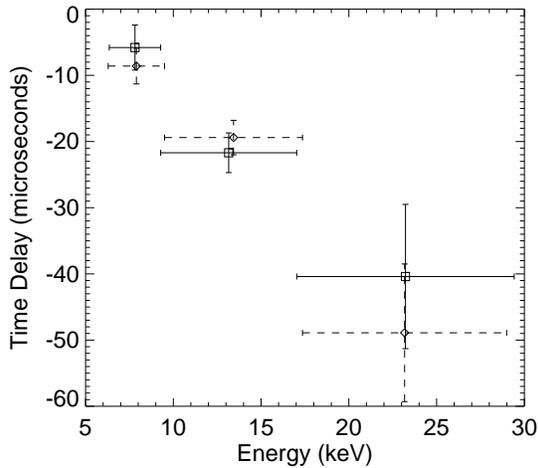}}
\caption{ Time delays of soft behind hard photons.}
\label{fig:3}
\source{Kaaret et al. 1999}
\end{figure}

\subsection{Saturation of the Frequency}

Several of the LMXB go through strong quasi-regular long time-scale
modulations. For 4U~1820--30 and 4U~1705--44 the time scale is about
half a year and the modulations are factors of 5 and 10,
respectively. For the transients 4U~1608--52 and Aql~X--1 the
amplitude is higher and the time scale longer (1-2 yr). For
4U~1636--54 and 4U~1728--34 the amplitude is a factor of 2 and the
time scale 30 days. Figure 4 (top) shows some {\em RXTE} ASM data for
4U~1820--30 \citep{Bloser00}.

\begin{figure}

\resizebox{0.95\columnwidth}{!}{
  \begin{tabular}{l}
  \epsfig{file=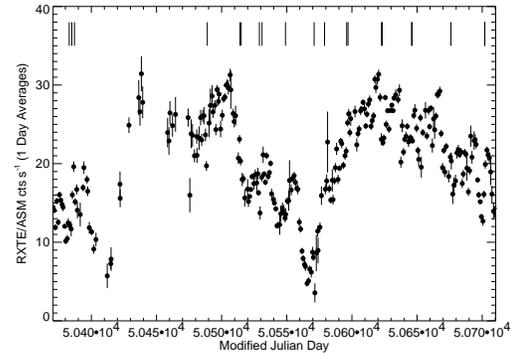,width=3.2in}\\
  \epsfig{file=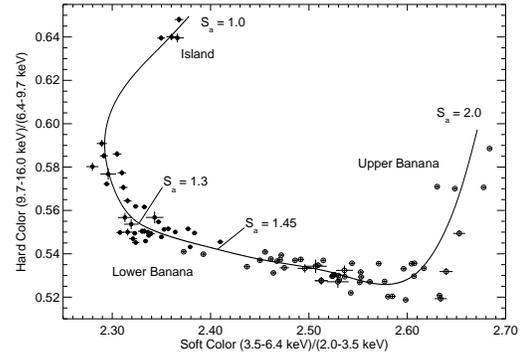,width=3.2in}\\
  \epsfig{file=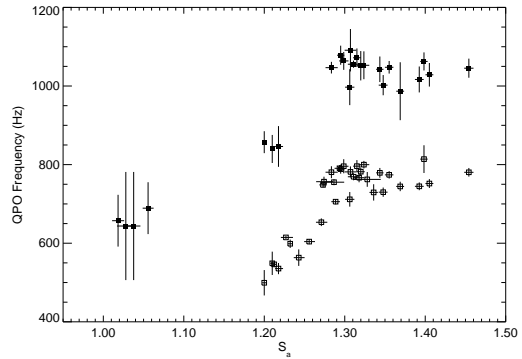,width=3.2in}
  \end{tabular}
}

\caption{4U 1820-30. (top) {\em RXTE} ASM light curve; (middle) Color-color diagram;
(bottom) QPO kilohertz frequencies versus S$_a$.}
\label{fig:4}
\source{Bloser et al. 2000} 
\end{figure}

For 4U~1820--30 the QPO frequencies increase as the source goes from
the ``Island'' to ``Lower Banana'' part of the color-color diagram
(CCD) shown in Figure 4 (middle).  In the ``Upper Banana'' part of the
diagram, QPO cannot be detected. The parameter S$_a$ tracks position
in the CCD and the increase of QPO frequency. For 4U~1820--30, the
$\geq$ 2 keV flux and luminosity approximately track S$_a$ as well, while in
other sources, the situation is more
complicated. But in 4U~1820--30, as shown in Figure 4 (bottom), the
frequency of both of the kilohertz oscillations increases with S$_a$
to a certain point and while S$_a$ increases further, the QPO
frequency has saturated. Because in this case S$_a$ is correlated with
luminosity, the source behaves as if the accretion rate increases with
the luminosity and pushes the inner disk inwards, with corresponding increases
in the Keplerian frequency at the inner edge of the disk.
%Once it reaches the ISCO the disk
%could not exist inside of it, so no higher frequencies would be
%possible. It should be possible to continue to have higher flux moving
%through the disk. \citet{Zhang98} first found that the frequency
%saturates. It was confirmed by \citet{Kaaret2Zhang99} in a different data
%set.  \citet{Bloser00} has changed the plot abscissa from luminosity to S$_a$.

\begin{figure}
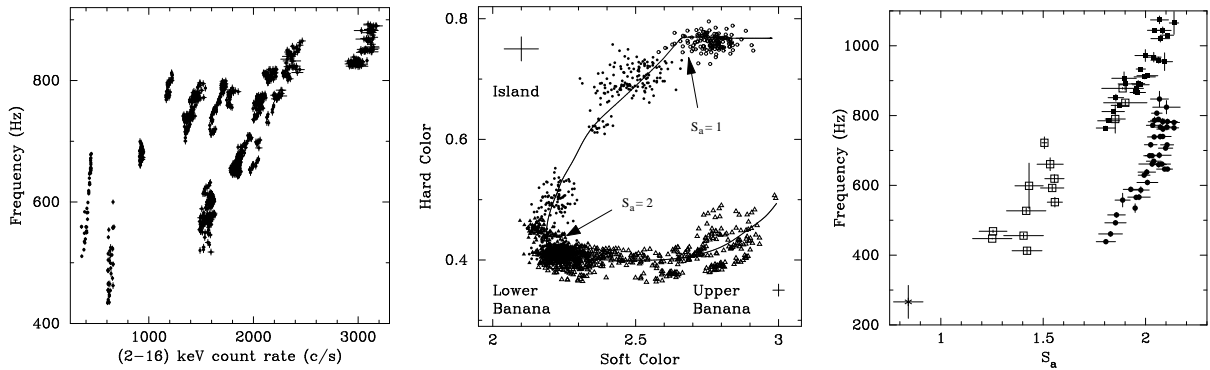


\resizebox{1.0\textwidth}{!}{
  \begin{tabular}{lcr}
  \epsfig{file=swankj_f5a.ps,width=2.0in,angle=-90,origin=c} & 
  \epsfig{file=swankj_f5b.ps,width=2.0in,angle=-90,origin=c} &
   \epsfig{file=swankj_f5c.ps,width=2.0in,angle=-90,origin=c} \\

  \end{tabular}
}

\caption{4U 1608--522: (left)Parallel tracks of frequency versus 
luminosity. (middle) Color-color diagram with assignment of scale of 
S$_a$. Each point represents 128 s of data, circles in the Island state, triangles in the Banana state, as defined by low frequency QPO properties, filled symbols exhibiting high frequency QPO, open ones no detected QPO. (right) Upper and lower kilohertz frequencies as a function of S$_a$, filled symbols for 
observations in which both QPO are present, the open squares when only one is
detected.
}
\label{fig:5}
\source{M\'{e}ndez et al. 1999} 

\end{figure}

Once it reaches the ISCO the disk
could not exist inside of it, so no higher frequencies would be
possible. It should be possible to continue to have higher flux moving
through the disk. \citet{Zhang98} first found that the frequency
saturates. It was confirmed by \citet{Kaaret2Zhang99} in a different data
set.  \citet{Bloser00} has changed the plot abscissa from luminosity to S$_a$.

\subsection{Unification of Atoll Source Parallel Tracks}

One of the confusing aspects of the kilohertz oscillations has been
that frequency was correlated with luminosity locally in time, but not
over long time scales \citep{Mendez99, Zhang2Zhang98}.  
As shown in Figure 5, multiple X-ray
luminosities have the same frequency.  In fact these tracks map onto a
color- color diagram with an S$_a$ defined such that the frequency is
a monotonic function of S$_a$. The frequency and the spectrum are
determined by S$_a$, while the luminosity is not.  It is
supposed that  S$_a$ represents a relevant accretion rate.

%\subsubsection{Models of the Parallel Tracks}

While the X-ray luminosity is apparently not the independent variable
determining the frequency and other properties of the QPO, the system
is simple enough that there is one controlling independent
variable. S$_a$ is unsatisfactory in not yet being
understood in terms of physical quantities. But it does determine the
properties. \citet{Mendez01} have shown that the QPO of 3 Atoll
bursters have almost identical energy spectra and runs of rms
amplitude as a function of the centroid frequency.

They have examined examples of multiple tracks and shown that for QPO
on different tracks, when the frequency is the same, the rms amplitude
is nearly the same.  It is not the case that an extra source of
luminosity varies that does not participate in the oscillation. A
suggested
simple picture was that there were two different flows, one
through the disk, one a radial or coronal flow. This model
is clearly ruled out.

A phenomenological model that generates multiple tracks was
constructed by \citet{vdK01}. He assumed there are two independent
time scales, one on the order of hours on which frequency and
luminosity is correlated, and one on the other of days or weeks, which
changes the scale of the luminosity. Thus the frequency is correlated,
not to the luminosity or an accretion rate directly, but to a form of
a fractional accretion rate. The whole surface of an accreting disk
can contribute to evaporation into a radial flow, while fluctuations
in the disk must diffuse through the disk. A quantitative physical
model has not been worked out, but some general predictions are made which 
could be borne out by persistent observations. M\'{e}ndez is exploring 
the characteristics of transitions, which should be sequential in that model. 
%It has been difficult to 
%catch the transitions \citep{M\'{e}ndez03}.

The spectrum determines the place in the color-color diagram. The
frequency is determined by that as well. These both could depend on
the radius of a ring where oscillations occur. Different luminosities
may correspond to different rates of flow through this region. Van der
Klis hypothesizes that a range of local equilibria of radiation
pressure, ram pressure and magnetic pressure are possible which allow 
the spectrum and oscillation forming conditions to be the same.

Long term modulations have in several sources (Her X--1, LMC X--4, SMC
X--1, and SS433) appeared to be due to shadowing by a tilted and
precessing accretion disk. Energy independent obscuration might be
able to explain aspects of the multiple tracks, but the time scales
for the long term changes are generally very irregular, if not
chaotic.  The simulations carried out by  \citet{vdK01}, which show promise of
representing important aspects of the behavior, assume a random walk in
the disk accretion rate. Further, some long-term variations are the 
same as motion in the Atoll diagram. 
%The spatial scales of the disks are quite different.

\begin{figure}[h]
 \resizebox{.8\columnwidth}{!}
%  {\includegraphics{../figs/homans_gx17+2.ps}}
  {\includegraphics{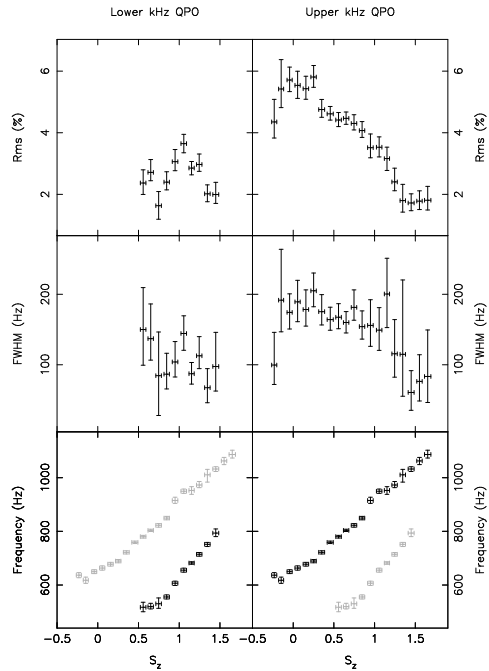}}
 \caption{Properties of the kilohertz QPO in the Z source GX 17+2,
 (left) $\nu 1$ and (right) $\nu 2$. S$_Z$ is defined along 
 the color-color diagram, such as to have the value 1.0 at the Horizontal
 Branch to Normal Branch inflection and the value 2.0 at the Normal 
 Branch to Flaring Branch transition.}
 \label{fig:6}
 \source{Homan et al. 2002} 
\end{figure}

\subsection{Z Sources}

While there have been suggestions of how Atoll and Z sources may be
related, the color-color diagrams and the ranges of luminosity
differ. The Z sources  have their own family
characteristics. 
%Figure 6 shows how they
%looked in early observations by the {\em RXTE} PCA.  
Ever since 
observations of Cyg X--2 showed changes in X-ray luminosity
anticorrelated with UV flux \citep{Has90,Vrtilek90},  we have
had to live with the idea that  the accretion
rate appears to increase along the Z, although the observed X-ray luminosity
increases along the ``Horizontal Branch'', decreases along the ``Normal
Branch'' and increases again in the ``Flaring Branch''.
In Figure 6 \citet{Homan02} shows typical dependences 
of the two kilohertz QPO properties
on the S$_Z$ parameter as it moves along the ``Z'' for GX~17+2. The
shape of the color versus intensity diagram that inspired the name
``Z'' depends on the instrumental definition of the color. As in the
case of the Atoll sources, some variable determines the spectrum and
the frequencies, a parameter with which the X-ray luminosity is
locally correlated or anticorrelated.

\subsection{Summary Parameters}

In general the QPO have higher amplitude at higher energies (at least
to about 20 keV) both in the case of the Atoll and the Z sources. The
amplitudes are higher for the Atoll sources than for the Z sources.
Although not invariably, there is a tendency for the
lower of the two kilohertz oscillations to be stronger and narrower
than the upper one. Table 2 shows the range of properties.

\begin{table}[h]
\begin{tabular}{lrr}
\hline 
& \tablehead{1}{l}{b}{Atoll Source} &
  \tablehead{1}{l}{b}{Z Source}\\
\hline
rms (E$\leq$6 keV) & 15 \% & 5 \%\\
rms (E $>$ 6 keV)  & 40 \% & 13 \%\\ 
$\Delta \nu 1$ & 5-100 Hz  & 70-200 Hz \\
$\Delta \nu 2$ & 10-200 HZ & 60-300 Hz \\ 
\hline
\end{tabular}
\caption{Amplitudes and  Widths of the Two Kilohertz QPO.}
\label{tab:b}
\end{table}

\subsection{Relation of the Difference Frequency to the Neutron Star Spin}

The {\em RXTE} observation of SAX J1808.4-3658 in October 2002 confirmed the
evidence of the BeppoSax observation \citep{in'tZand01} that bursts
from this source can exhibit pulsations. {\em RXTE} measured the frequency and
amplitude of the burst oscillations accurately. In addition  two 
kilohertz QPO were detected during a part of the outburst. 
It was not known whether these QPO would be
possible under the flow conditions that allow the coherent
pulsations to be seen. We now know that they can coexist under some
conditions. \citet{Wijnands03} found that two QPO appeared near the
peak of the outburst. As the outburst luminosity declined a single,
broad, QPO persisted and drifted lower in frequency, moving through
the pulse frequency with little apparent affect, as shown in Figure
7. When two QPO were present, the difference was consistent with being
half the spin frequency. SAX J1808.4-3658 thus has shown  that the
spin frequency is the frequency seen in the bursts, while the
difference can be about half that. A strong upper limit was established
on power at the 401 Hz sub-harmonic.

\begin{figure}
\resizebox{0.95\columnwidth}{!}
%{\includegraphics{../figs/saxj1808qpo.ps}} 
{\includegraphics{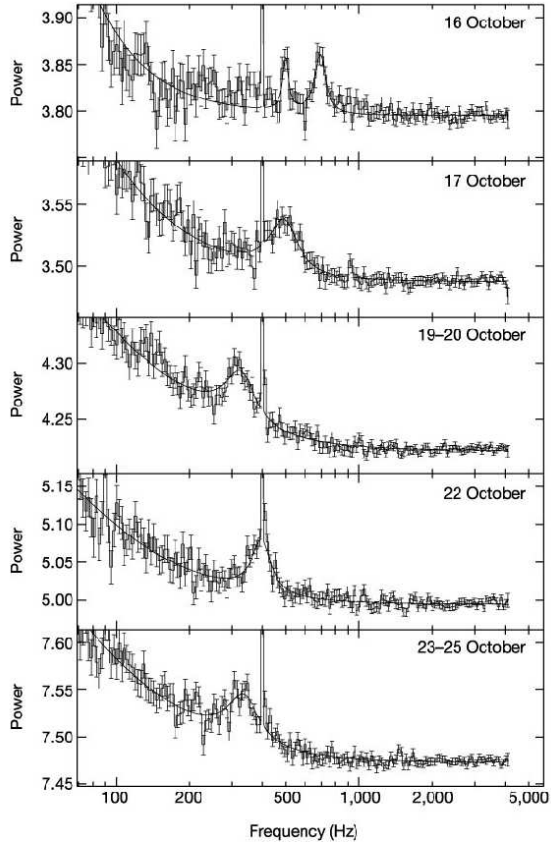}} 
\caption{SAX J1808.4-3658 power spectra}
\label{fig:7}
\source{Wijnands et al. 2003}
\end{figure}

%\subsection{Variations in $\Delta \nu$}

To first approximation for a given source in which two kilohertz
oscillations have been seen, the difference is approximately
constant. The data are accurate enough that to the next approximation
the differences can be seen to vary, in particular as functions of one
of the two frequencies. While the differences are as low as 1\% and
seldom larger than 15 \%, they are significant and systematic. In the
case of the relativistic precession model, the difference is specified
in terms the central mass and spin, and the radius, as is the
azimuthal frequency. It is possible to find fits ``in the ball park''
of what is required, although the fits are not statistically
acceptable. However it is not at all clear whether any
physical mechanism can give rise to x-ray emission modulated with the 
frequencies characteristic of eccentric particle orbits. 

The Atoll source difference
frequencies are plotted in Figure 8 against the lower kilohertz
frequency, along with the observed values of Sco X-1. We do not know
at this point, whether the fact that the difference frequencies bear
such a close relation to the neutron star spins means that they are
physically related, or whether the closeness can be coincidental. The
Sonic Point Beat Frequency Model appeared to imply that the difference
frequency should be the spin. That is now known to be incorrect,
although explanations are being explored.  Perhaps there is a
resonance which drives the spin to stabilize at a multiple of a
specific $\Delta \nu$.

\begin{figure}[t]
\resizebox{1.0\columnwidth}{!}
%{\includegraphics{../figs/atoll_ratio_grab.ps}}
{\includegraphics{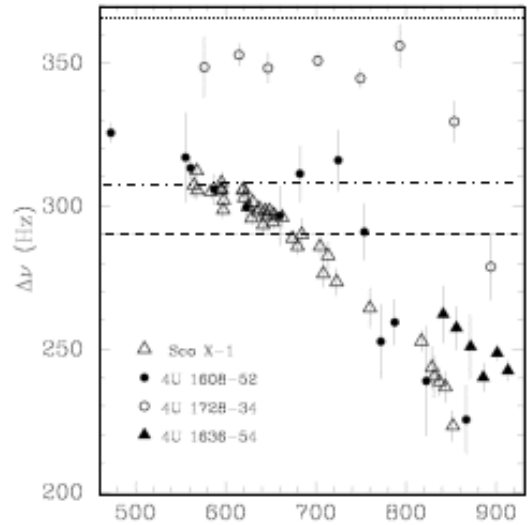}}
\caption{$\Delta \nu$ for Atoll Sources, with comparison to Sco X--1. 
The dashed line is 1/2 the 
spin frequency for 4U~1636--536. The dot-dash line is 1/2 the spin
frequency for 4U~1608--522. The dotted line is the spin frequency 
of 4U~1728--34.}
\label{fig:8}
\source{adapted from Di Salvo et al. 2003}
\end{figure}

\section{Low Frequency Oscillations}

Low frequency QPO (below about 50 Hz) are present in various 
forms in LMXBs with neutron
stars and in systems with black holes. They exhibit a variety of
strengths and coherence ($\nu/\Delta \nu$). Some
characteristics, notably band-limited white noise with a break frequency,
and
a QPO near the break frequency and closely correlated with it, appear
in Atoll sources in the Island state, in Z sources on the Horizontal
Branch, and in black hole sources in ``low hard'' and ``intermediate'' 
states.
%In some cases only a single broad low frequency
%QPO appears. 
In some cases this QPO is strong and a first harmonic is
observed. Sometimes a weak sub-harmonic appears
(e.g.\citep{JonkerGX5, JonkerGX340}).

The strongest low frequency QPO is notable for being correlated with the
kilohertz QPO.
In particular, it sometimes is proportional to the square of the 
upper kilohertz frequency \citep{vdK00}. This correlation in the case of 
GX~17+2 has been shown to reverse at the highest frequencies \citep{Homan02}.
{\em RXTE} observations have shown that the Normal 
Branch oscillations of Z sources can also
move in frequency with the kilohertz oscillations \citep{vdK00,JonkerGX5}.

It has been
pointed out that with certain identifications of features in the power
spectra of black hole candidates, the correlations appear to be the
same for neutron stars and black holes. It has recently been claimed
that even the the white dwarf SS Cygni exhibits oscillations that fall
on the same line.  The similarity of correlations clearly seems to
imply that their explanations are related. At first this may seem to
imply that neutron star characteristics, surface and spin, can not
be playing a role.  If it holds for a white dwarf system, it could
imply that the relations come from disk characteristics that are not
related to General Relativity.  It is likely that phenomena occur in a
disk because of properties that do not depend on the nature of the
compact object, but whose exact values do depend on how strong gravity
is  and what the boundary conditions are.  A careful evaluation
is needed of how exact and general are the correlations.

\section{Conclusions}

In the complex of QPO observed in data from the
LMXB containing neutron stars {\em RXTE} has found very clearly
signatures which are generally characteristic of the
sources. The signatures are related to those seen in black hole 
sources and perhaps to
phenomena seen in a few white dwarf systems. But the neutron star
systems have exhibited definitive characteristics that allow very
quantitative study of the dependences on parameters. Timing is a tool
for obtaining spatial resolution and understanding of dynamics. Even
without final detailed interpretation, the QPO are clearly
entirely consistent with the neutron stars being inside their
ISCOs and with the accretion disks penetrating so close to the neutron
star as to be affected by General Relativity. The set of signatures is
very simple compared to, say, atomic spectra, but intricate enough that the
correct explanation of the details will be interesting and important.

%%%%%%%%%%%%%%%%%%%%%%%%%%%%%%%%%%%%%%%%%%%%%%%%
%% BACKMATTER
%%%%%%%%%%%%%%%%%%%%%%%%%%%%%%%%%%%%%%%%%%%%%%%%

\begin{theacknowledgments}
  I  thank the many {\em RXTE} users whose results I discuss
here, the {\em RXTE} Science Operations Center members for the roles they
play in making the observations successful, and NASA for its support of 
all aspects of {\em RXTE}.
\end{theacknowledgments}

%%%%%%%%%%%%%%%%%%%%%%%%%%%%%%%%%%%%%%%%%%%%%%%%
%% You may have to change the BibTeX style below, depending on your
%% setup or preferences.
%%
%% If the bibliography is produced without BibTeX comment out the
%% following lines and see the aipguide.pdf for further information.
%%
%% For The AIP proceedings layouts use either
%%%%%%%%%%%%%%%%%%%%%%%%%%%%%%%%%%%%%%%%%%%%

\bibliographystyle{aipproc}   % if natbib is available
%\bibliographystyle{aipprocl} % if natbib is missing

%%%%%%%%%%%%%%%%%%%%%%%%%%%%%%%%%%%%%%%%%%%
%% You probably want to use your own bibtex database here
%%%%%%%%%%%%%%%%%%%%%%%%%%%%%%%%%%%%%%%%%%%
\bibliography{swankj}

\hyphenation{Post-Script Sprin-ger}
\begin{thebibliography}{34}
\expandafter\ifx\csname natexlab\endcsname\relax\def\natexlab#1{#1}\fi
\providecommand{\enquote}[1]{``#1''}
\expandafter\ifx\csname url\endcsname\relax
  \def\url#1{\texttt{#1}}\fi
\expandafter\ifx\csname urlprefix\endcsname\relax\def\urlprefix{URL }\fi

\bibitem[Alpar et~al.(1982)]{alpar82}
Alpar, M.~A., Cheng, A.~F., Ruderman, M.~A., and Shaham, J., \emph{Nature},
  \textbf{300}, 23--30 (1982).

\bibitem[Hasinger and van~der Klis(1989)]{HasvdK89}
Hasinger, G., and van~der Klis, M., \emph{Astronomy and Astrophysics},
  \textbf{225}, 79--96 (1989).

\bibitem[Wijnands(2001)]{Wijnands01}
Wijnands, R., \emph{Advances in Space Research}, \textbf{28}, 469--479 (2001).

\bibitem[Alpar and Shaham(1985)]{alpar85}
Alpar, M.~A., and Shaham, J., \emph{Nature}, \textbf{316}, 239--241 (1985).

\bibitem[Ghosh and Lamb(1992)]{GL92}
Ghosh, P., and Lamb, F.~K., \enquote{Diagnostics of disk-magnetosphere
  interaction,} in \emph{X-Ray Binaries and Recycled Pulsars}, edited by
  E.~van~den Heuvel and S.~A. Rappaport, NATO ASI Series C 377, Kluwer Academic
  Publishers, Boston, 1992, pp. 487--510.

\bibitem[van~der Klis et~al.(1990)]{vdK90}
van~der Klis, M., Hasinger, G., Damen, E., Penninx, W., van Paradijs, J., and
  Lewin, W. H.~G., \emph{Astrophysical Journal}, \textbf{360}, L19--L22 (1990).

\bibitem[Hasinger et~al.(1990)]{Has90}
Hasinger, G., van~der Klis, M., Ebisawa, K., Dotani, T., and Mitsuda, K.,
  \emph{Astronomy and Astrophysics}, \textbf{235}, 131--146 (1990).

\bibitem[Vrtilek et~al.(1990)]{Vrtilek90}
Vrtilek, S.~D., Raymond, J.~C., Garcia, M.~R., Verbunt, F., Hasinger, G., and
  Kurster, M., \emph{Astronomy and Astrophysics}, \textbf{235}, 162--173
  (1990).

\bibitem[Vrtilek et~al.(1991)]{Vrtilek91}
Vrtilek, S., Penninx, W., Raymond, J., Verbunt, F., Hertz, P., Wood, K., Lewin,
  W. H.~G., and Mitsuda, K., \emph{Astrophysical Journal}, \textbf{376},
  278--288 (1991).

\bibitem[Vaughan(1994)]{Vaughan94}
Vaughan, e.~a., B.~A., \emph{Astrophysical Journal}, \textbf{435}, 362--371
  (1994).

\bibitem[Kluzniak and Wagoner(1985)]{Kluz85}
Kluzniak, W., and Wagoner, R.~V., \emph{Astrophysical Journal}, \textbf{297},
  548--554 (1985).

\bibitem[Kluzniak et~al.(1990)]{Kluz90}
Kluzniak, W., Michelson, P., and Wagoner, R.~V., \emph{Astrophysical Journal},
  \textbf{358}, 538--544 (1990).

\bibitem[Strohmayer et~al.(1996)]{Stroh96}
Strohmayer, T.~E., Zhang, W., Swank, J.~H., Smale, A., Titarchuk, L., Day, C.,
  and Lee, U., \emph{Astrophysical Journal}, \textbf{469}, L9--L12 (1996).

\bibitem[van~der Klis et~al.(1996)]{vdK96}
van~der Klis, M., Swank, J.~H., Zhang, W., Jahoda, K., Morgan, E.~H., Lewin, W.
  H.~G., Vaughan, B., and van Paradijs, J., \emph{Astrophysical Journal},
  \textbf{469}, L1--L4 (1996).

\bibitem[van~der Klis(2000)]{vdK00}
van~der Klis, M., \emph{Annual Review of Astronomy and Astrophysics},
  \textbf{38}, 717--760 (2000).

\bibitem[Strohmayer et~al.(1998)]{SSZ98}
Strohmayer, T.~E., Swank, J.~H., and Zhang, W., \emph{Nuclear Physics B (Proc
  Suppl.)}, \textbf{69}, 129--134 (1998).

\bibitem[van Straaten et~al.(2000)]{Straaten00}
van Straaten, S., Ford, E.~C., van~der Klis, M., M\'{e}ndez, M., and Kaaret,
  P., \emph{Astrophysical Journal}, \textbf{540}, 1049--1061 (2000).

\bibitem[Zhang et~al.(1997)]{Zhang97}
Zhang, W., Strohmayer, T.~E., and Swank, J.~H., \emph{Astrophysical Journal},
  \textbf{482}, L167--L170 (1997).

\bibitem[Kaaret et~al.(1997)]{Kaaret97}
Kaaret, P., Ford, E.~C., and Chen, K., \emph{Astrophysical Journal},
  \textbf{480}, L27--L29 (1997).

\bibitem[Ford et~al.(2000)]{Ford00}
Ford, E.~C., van~der Klis, M., M\'{e}ndez, M., Wijnands, R., Homan, J., Jonker,
  P.~G., and van Paradijs, J., \emph{Astrophysical Journal}, \textbf{537},
  368--373 (2000).

\bibitem[White and Zhang(1997)]{White97}
White, N., and Zhang, W., \emph{Astrophysical Journal}, \textbf{490}, L87--L90
  (1997).

\bibitem[Kaaret et~al.(1999{\natexlab{a}})]{Kaaret99}
Kaaret, P., Piraino, S., Ford, E.~C., and Santangelo, A., \emph{Astrophysical
  Journal}, \textbf{514}, L31--L33 (1999{\natexlab{a}}).

\bibitem[Bloser et~al.(2000)]{Bloser00}
Bloser, P.~F., Grindlay, J.~E., Kaaret, P., Zhang, W., Smale, A.~P., and
  Barret, D., \emph{Astrophysical Journal}, \textbf{542}, 1000--1015 (2000).

\bibitem[Zhang et~al.(1998{\natexlab{a}})]{Zhang98}
Zhang, W., Smale, A., Strohmayer, T.~E., and Swank, J.~H., \emph{Astrophysical
  Journal}, \textbf{500}, L171--L174 (1998{\natexlab{a}}).

\bibitem[Kaaret et~al.(1999{\natexlab{b}})]{Kaaret2Zhang99}
Kaaret, P., Piraino, S., Bloser, P.~F., Ford, E.~C., Grindlay, J.~E.,
  Santangelo, A., and Zhang, W., \emph{Astrophysical Journal}, \textbf{520},
  L37--L40 (1999{\natexlab{b}}).

\bibitem[M\'{e}ndez et~al.(1999)]{Mendez99}
M\'{e}ndez, M., van~der Klis, M., Ford, E.~C., Wijnands, R., and van Paradijs,
  J., \emph{Astrophysical Journal}, \textbf{511}, L49--L52 (1999).

\bibitem[Zhang et~al.(1998{\natexlab{b}})]{Zhang2Zhang98}
Zhang, W., Jahoda, K., Kelley, R.~L., Strohmayer, T.~E., Swank, J.~H., and
  Zhang, S.~N., \emph{Astrophysical Journal}, \textbf{495}, L9--L12
  (1998{\natexlab{b}}).

\bibitem[M\'{e}ndez et~al.(2001)]{Mendez01}
M\'{e}ndez, M., van~der Klis, M., and Ford, E.~C., \emph{Astrophysical
  Journal}, \textbf{561}, 1016--1026 (2001).

\bibitem[van~der Klis(2001)]{vdK01}
van~der Klis, M., \emph{Astrophysical Journal}, \textbf{561}, 943--949 (2001).

\bibitem[Homan et~al.(2001)]{Homan02}
Homan, J., van~der Klis, M., Jonker, P.~G., Wijnands, R., Kuulkers, E.,
  M\'{e}ndez, M., and Lewin, W. H.~G., \emph{Astrophysical Journal},
  \textbf{568}, 878--900 (2001).

\bibitem[in't Zand et~al.(2001)]{in'tZand01}
in't Zand, J.~J., Cornelisse, R., Kuulkers, E., Heise, J., Kuiper, L., Bazzano,
  A., Cocci, M., Muller, J.~M., Natalucci, L., Smith, M. J.~S., and Ubertini,
  P., \emph{Astronomy and Astrophysics}, \textbf{372}, 916--921 (2001).

\bibitem[Wijnands et~al.(2003)]{Wijnands03}
Wijnands, R., van~der Klis, M., Homan, J., Chakrabarty, D., Markwardt, C.~B.,
  and Morgan, E.~H., \emph{Nature}, \textbf{424}, 44--47 (2003).

\bibitem[Jonker et~al.(2002)]{JonkerGX5}
Jonker, P.~G., van~der Klis, M., Homan, J., M\'{e}ndez, M., Lewin, W. H.~G.,
  Wijnands, R., and Zhang, W., \emph{MNRAS}, \textbf{333}, 665--678 (2002).

\bibitem[Jonker et~al.(2000)]{JonkerGX340}
Jonker, P.~G., van~der Klis, M., Wijnands, R., Homan, J., van Paradijs, J.,
  M\'{e}ndez, M., Ford, E.~C., Kuulkers, E., and Lamb, F.~K.,
  \emph{Astrophysical Journal}, \textbf{537}, 374--386 (2000).

\end{thebibliography}

%%%%%%%%%%%%%%%%%%%%%%%%%%%%%%%%%%%%%%%%%%%
%% Just a reminder that you may have to run bibtex
%% All of it up to \end{document} can be removed
%% if you don't like the warning.
%%%%%%%%%%%%%%%%%%%%%%%%%%%%%%%%%%%%%%%%%%%
\IfFileExists{\jobname.bbl}{}
 {\typeout{}
  \typeout{******************************************}
  \typeout{** Please run "bibtex \jobname" to optain}
  \typeout{** the bibliography and then re-run LaTeX}
  \typeout{** twice to fix the references!}
  \typeout{******************************************}
  \typeout{}
 }

\end{document}